\documentclass[prx,onecolumn,english,superscriptaddress,floatfix,longbibliography]{revtex4-2}

\usepackage[english]{babel}
\usepackage{amsmath}
\usepackage{mhchem}
\usepackage{tabularx}
\usepackage{adjustbox}
\usepackage{booktabs}
\usepackage[letterpaper,top=2cm,bottom=2cm,left=3cm,right=3cm,marginparwidth=1.75cm]{geometry}

\usepackage{amsmath}
\usepackage{graphicx}
\usepackage[colorlinks=true, allcolors=blue]{hyperref}
\usepackage{braket}
\usepackage{tikz}
\usepackage{xcolor}
\usepackage{subcaption}
\usepackage{amsthm}

\newtheorem{theorem}{Theorem}[section]
\usetikzlibrary{matrix}
\definecolor{lightred}{HTML}{ffcccb}
\definecolor{lightcyan}{HTML}{E0FFFF}
\definecolor{lightgray}{HTML}{f8f8f8}

\tikzset{square matrix/.style={
    matrix of nodes,
    column sep=-\pgflinewidth, row sep=-\pgflinewidth,
    nodes={draw=none,
      minimum height=#1,
      anchor=center,
      text width=#1,
      align=center,
      inner sep=0pt
    },
  },
  square matrix/.default=0.6cm
}

\begin{document}

\title{Pascal's pyramid and number projection operators for quantum computation}

\author{Dario Picozzi}

\affiliation{Department of Physics and Astronomy, University College London (UCL), Gower Street, London, WC1E 6BT, United Kingdom}

\email{picozzi.dario@gmail.com}

\maketitle

\section*{Abstract}

The pursuit of quantum advantage in simulating many-body quantum systems on quantum computers has gained momentum with advancements in quantum hardware. This work focuses on leveraging the symmetry properties of these systems, particularly particle number conservation. We investigate the qubit objects corresponding to number projection operators in the standard Jordan-Wigner fermion-to-qubit mapping, and prove a number of their properties. This reveals connections between these operators and the generalised binomial coefficients originally introduced by Kravchuk in his research on orthogonal polynomials. The generalized binomial coefficients are visualized in a Pascal's pyramid structure.

\section{Introduction}

As we are able to build better, more powerful quantum hardware, the goal of quantum advantage for useful tasks seems increasingly to be within reach. Over the last few decades, the simulation of many-body quantum systems on quantum computers and in particular molecular systems \cite{Whitfield2011, Cao2019, McArdle2020, Leontica2021} has been singled out as one of the areas where an advantage over classical computing might be found in the near-term.

Philip W. Anderson famously claimed that 'it is only slightly overstating the case to say that physics is the study of symmetry' \cite{Anderson_1972}: the correspondence between symmetries of the simulated systems in the physical word and the objects in the quantum computer is a fruitful field, and this understanding can provide benefits in terms of computational resources and accuracy.

A feature that emerges from the non-relativistic approximation of many-body quantum systems is particle number conservation symmetry. Several methods have been proposed to enforce number conservation and other symmetries on quantum computing simulations: these include symmetry projectors \cite{Yen_Lang_Izmaylov_2019} and symmetry-adapted encodings \cite{Picozzi_Tennyson_2023}.

In what follows we consider the qubit objects that correspond to number projection operators in the standard Jordan-Wigner fermion-to-qubit mapping \cite{Jordan1928}. We explore a number of properties and their relationship to the generalised binomial coefficients first introduced by Kravchuk \cite{Krawtchouk1929} in the study of orthogonal polynomials.

A straightforward application of the form of the number projection operators we describe allows a Jordan-Wigner qubit operator $\hat{M}$ to be converted into a projected qubit operator on the subspace of $k$ particles in $n$ spin-orbitals by conjugation $\hat{M}_P = \hat{P}\hat{M}\hat{P}$. A potential drawback of this is that the number of Pauli terms increases exponentially in the number of qubits, although we find that the number of Pauli clique partitions \cite{Jena_2019, Gokhale2020, Kurita2023} maintains a similar scaling with the number of qubits.

\begin{figure}
\centering
\includegraphics[width=0.6\linewidth]{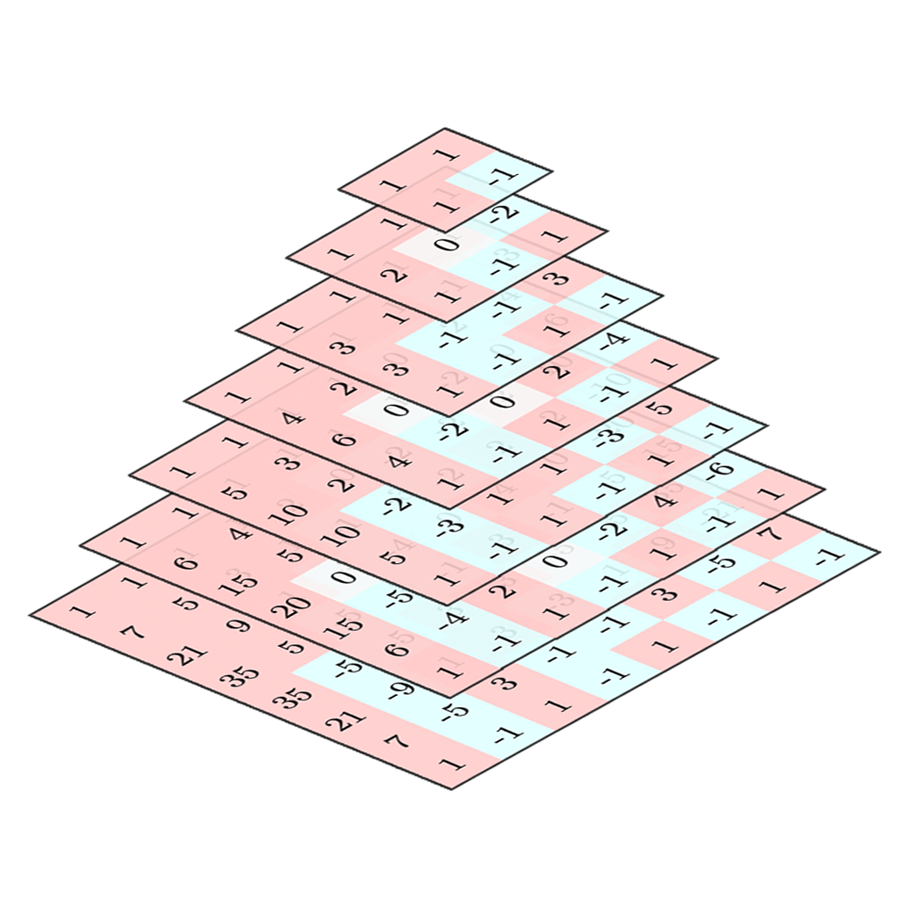}
\caption{Pascal's pyramid for the generalised binomial coefficients $C(n, k, m)$: the left-facing side of the pyramid corresponds to Pascal's triangle, where the downward direction corresponds to increasing $n$ index and the sideways direction to increasing $k$ index; the additional $m$ index is given by the extra dimension. Both $k$ and $m$ range from $0$ to $n$ (from left to right). The values of the generalised binomial coefficients $C(n, k m)$ for $1 \leq n \leq 8$ are shown fully in Fig. \ref{fig:generalised_binomial_coefficients_tables}}
\label{fig:pascals_pyramid}
\end{figure}

\section{Results}

The number projection operator $\hat{P}(n, k)$ is the qubit operator that maps computational basis states with $k$ out of $n$ bits equal to $1$ (i.e. whose Hamming weight is equal to $k$) to themselves, and that maps all other computational basis states to 0. For example:

\begin{equation}
\begin{split}
    \hat{P}(3, 1) \ket{010} & = \ket{010} \\
    \hat{P}(3, 1) \ket{011} & = 0
\end{split}
\end{equation}

The Pauli decomposition of the number projection operator $\hat{P}(n, k)$ is given by the weighted sum of Pauli $Z$ operators:
\begin{equation}
    \hat{P}(n, k) = \frac{1}{2^n} \sum_{m = 0}^{n}  {C(n, k, m) \sum_{\{i_1, \dots, i_m\} \in I_m} {Z_{i_1} \otimes \dots \otimes Z_{i_m}}}
\label{eq:Pauli_expansion_of_projector}
\end{equation}
Where $I_m$ is the set of the $\binom{n}{m}$ combinations of $m$ numbers from $0$ to $n - 1$.

For example for $n = 3$ and $k = 1$ we have $C(3, 1, 0) = 3$, $C(3, 1, 1) = 1$, $C(3, 1, 2) = -1$, and $C(3, 1, 3) = -3$:
\begin{equation}
\begin{split}
    \hat{P}(3, 1) &= \ket{001}\bra{001} + \ket{010}\bra{010} + \ket{100}\bra{100}\\
    &= \frac{3}{8} III + \frac{1}{8}(IIZ + IZI + ZII) - \frac{1}{8}(IZZ + ZIZ + ZZI) - \frac{3}{8}ZZZ
\end{split}
\end{equation}

\begin{theorem}[Generalised binomial coefficients]
The coefficient $C(n, k, m)$ for the tensor product of $m$ Pauli $Z$ operators in the expansion for the projector $n$ choose $k$ is given by:
\begin{equation}
    \boxed{C(n, k, m) = \sum_{l = 0}^m (-1)^l \binom{n - m}{k - l} \binom{m}{l}}
\label{eq:closed_form}
\end{equation}

In particular the coefficient $C(n, k, 0)$ for the identity term ($m = 0$) is simply the binomial coefficient $\binom{n}{k}$:

\begin{equation}
    C(n, k, 0) = \binom{n}{k}
\label{eq:binomial}
\end{equation} 

The coefficients $C(n, k, m)$ also satisfy the addition recursive relation (for $m \neq n$):
\begin{equation}
    \boxed{C(n, k, m) = C(n - 1, k, m) + C(n - 1, k - 1, m)}
\label{eq:recursive_plus}
\end{equation}

And the subtraction recursive relation (for $m \neq 0$):
\begin{equation}
    \boxed{C(n, k, m) = C(n - 1, k, m - 1) - C(n - 1, k - 1, m - 1)}
\label{eq:recursive_minus}
\end{equation}
\end{theorem}

\begin{proof}
The projector $\hat{P}(n, k)$ has the computational basis states as its eigenvectors, and hence it is diagonal in the computational basis: therefore its Pauli decomposition only contains tensor products of Pauli $Z$ operators and the identity. Furthermore, $\hat{P}(n, k)$ is invariant under a permutation of the qubits (it only counts if exactly $k$ out of $n$ bits are equal to 1, independently of their order): this allows us to define the coefficients $C(n, k, m)$ as in Eq. \ref{eq:Pauli_expansion_of_projector}.

In particular, the projector operators on one qubit (i.e. for $n = 1$) are:
\begin{equation}
\begin{split}
    \hat{P}(1, 0) & = \ket{0}\bra{0} = \frac{I + Z}{2} \\
    \hat{P}(1, 1) & = \ket{1}\bra{1} = \frac{I - Z}{2}
\end{split}
\label{eq:n=1}
\end{equation}
Which yields the coefficients $C(1, 0, 0) = C(1, 0, 1) = C(1, 1, 0) = 1$ and $C(1, 1, 1) = -1$.

Moreover, the projector $\hat{P}(n, k)$ can be written recursively in terms of the projectors $\hat{P}(n - 1, k)$ and $\hat{P}(n - 1, k - 1)$ as:
\begin{equation}
\begin{split}
    \hat{P}(n, k) & = \hat{P}(1, 0) \otimes \hat{P}(n - 1, k) + \hat{P}(1, 1) \otimes \hat{P}(n - 1, k - 1) \\
    & = \frac{I + Z}{2} \otimes \hat{P}(n - 1, k) + \frac{I - Z}{2} \otimes \hat{P}(n - 1, k - 1) \\
    & = I \otimes \frac{\hat{P}(n - 1, k) + \hat{P}(n - 1, k - 1)}{2} + Z \otimes \frac{\hat{P}(n - 1, k) - \hat{P}(n - 1, k - 1)}{2} 
\end{split}
\end{equation}
From the first term in the last line the coefficient $C(n, k, m)$ for a Pauli term that does not contain a Pauli $Z$ operator on the $n$-th (leftmost) qubit in the expansion for $\hat{P}(n, k)$ and that contains $m$ $Z$ operators overall must be equal to the sum of the coefficient $C(n - 1, k, m)$ and the coefficient $C(n - 1, k, m)$, which proves Eq. \ref{eq:recursive_plus}. From the second term the coefficient $C(n, k, m)$ for a Pauli term that contains a Pauli $Z$ operator on the $n$-th (leftmost) qubit in the expansion for $\hat{P}(n, k)$ and that contains $m$ $Z$ operators overall must be equal to the difference of the coefficient $C(n - 1, k, m - 1)$ and the coefficient $C(n - 1, k, m - 1)$, which proves Eq. \ref{eq:recursive_minus}.

Eq. \ref{eq:binomial} ($m = 0$) follows by induction from the recursive relation of Eq. \ref{eq:recursive_plus} together with the base case $n = 1$ in Eq. \ref{eq:n=1}.

Eq. \ref{eq:closed_form} then follows by induction from Eq. \ref{eq:binomial}:
\begin{equation}
\begin{split}
    C(n, k, m) &= \sum_{l = 0}^m (-1)^l \binom{n - m}{k - l} \binom{m}{l}\\
    &= \sum_{l = 0}^{m} (-1)^l \binom{n - m}{k - l} \left( \binom{m - 1}{l} + \binom{m - 1}{l-1} \right)\\
    &= \sum_{l = 0}^{m-1} (-1)^l \binom{n - m}{k - l} \binom{m - 1}{l} + \sum_{l = 1}^{m} (-1)^{l} \binom{n - m}{k - l} \binom{m - 1}{l-1}\\
    &= \sum_{l = 0}^{m-1} (-1)^l \binom{n - m}{k - l} \binom{m - 1}{l} - \sum_{l = 0}^{m-1} (-1)^l \binom{n - m}{k - l - 1} \binom{m - 1}{l}\\
    &= C(n - 1, k, m - 1) - C(n - 1, k - 1, m - 1)
\end{split}
\end{equation}
\end{proof}

The generalised coefficients $C(n, k, m)$ are often referred to in the literature as the Kravchuk matrices (or Krawtchouk matrices) \cite{Krawtchouk1929}. These have applications in harmonic analysis \cite{Szego1975}, combinatorics \cite{MacWilliams1963, Levenstein1995}, statistics and probability \cite{Feinsilver1991}, and coding theory \cite{Delsarte1972}. More recently, they have appeared in the theory of quantum walks \cite{Feinsilver2004} and in quantum gate architecture\cite{Groenland_Schoutens_2018}. In more theoretical developments, their relationship to the representation theory of the rotation groups has been investigated \cite{Genest2013}. They are most straightforwardly defined as the generalization of the binomial coefficients such that:

\begin{equation}
    (1 - x)^{m}(1 + x)^{n - m} = \sum_{k = 0}^{n} {C(n, k, m)x^{k}}
\end{equation}

The relationship with the traditional binomial coefficients is visualised in Pascal's pyramid in Fig \ref{fig:pascals_pyramid}. The values of the generalised binomial coefficients $C(n, k m)$ for $1 \leq n \leq 8$ are shown fully in Fig. \ref{fig:generalised_binomial_coefficients_tables}.

In the original formulation, the generalised binomial coefficient $C(n, k, m)$ is the value of the $k$-th Kravchuk polynomial with parameter $n$ evaluated at $x = m$, the set of polynomials that are orthogonal with respect to the binomial distribution. 

In what follows we derive a number of properties of the generalised binomial coefficients, using the definition in terms of the qubit projection operators in Eq. \ref{eq:Pauli_expansion_of_projector} as the starting point.

\begin{theorem}[Sum over columns]
\begin{equation}
    \sum_{k = 0}^n C(n, k, m) =     \begin{cases} 
    2^n \quad \text{for } m = 0\\
    0 \quad \text{for } m \neq 0
   \end{cases}
\end{equation}
\end{theorem}
\begin{proof}
The sum of projection operators $\hat{P}(n, k)$ over all values $0 \leq k \leq n$ is the identity:
\begin{equation}
    \sum_{k = 0}^n \hat{P}(n, k) = 1
\end{equation}
Comparing term by term we obtain the result.
\end{proof}

\begin{theorem}[Row orthogonality]
\begin{equation}
    \sum_{m = 0}^n \binom{n}{m} C(n, k, m) C(n, k', m) =     \begin{cases} 
    2^n \quad \text{for } k = k'\\
    0 \quad \text{otherwise}
   \end{cases}
\end{equation}
\end{theorem}
\begin{proof}
Applying a projection twice is the same as applying it once ($\hat{P}^2 = \hat{P}$) and the product of projections $\hat{P}(n, k)$ and $\hat{P}(n, k')$ over a different number of qubits ($k \neq k'$) is the zero operator:
\begin{equation}
    \hat{P}(n, k) \hat{P}(n, k') =     \begin{cases} 
      \hat{P}(n, k) & \quad \text{if } k = k' \\
      0 & \quad \text{otherwise} \\
   \end{cases}
\end{equation}
Comparing Pauli decompositions for both expressions and considering the coefficient for the identity term on both sides we obtain the result. In particular, for the LHS only terms that are squares of the same Pauli tensor contribute to the identity, of which there are in general $2^n$, and where the coefficient $C(n, k, m)$ appears $\binom{n}{m}$ times.

(Remark: this theorem is equivalent to the statement of the orthogonality of the Kravchuk polynomials with respect to the binomial distribution)
\end{proof}

\begin{theorem}[Sum over rows]
\begin{equation}
    \sum_{m = 0}^n \binom{n}{m} C(n, k, m) =     \begin{cases} 
    2^n \quad \text{for } k = 0\\
    0 \quad \text{for } k \neq 0
   \end{cases}
\end{equation}
\end{theorem}
\begin{proof}
From the previous result with $k' = 0$.
\end{proof}

\begin{theorem}[Sum over columns with number operator]
\begin{equation}
    \sum_{k = 0}^n k C(n, k, m) =
    \begin{cases} 
      n 2^{n-1} & \quad \text{for } m = 0 \\
      -2^{n-1} & \quad \text{for } m = 1 \\
      0 & \quad \text{for } 1 < m \leq n 
   \end{cases}
\label{eq:number_operator_identity}
\end{equation}
\end{theorem}
\begin{proof}
The number operator can be expressed in terms of the projection operators on $n$ qubits as the sum:
\begin{equation}
    \hat{N} = \sum_{k = 0}^n k \hat{P}(n, k)
\end{equation}
Its Pauli decomposition is however given by:
\begin{equation}
    \hat{N} = \frac{1}{2} (n - \sum_{l = 0}^{n-1}{Z_l})
\end{equation}
Comparing term by term we obtain Eq \ref{eq:number_operator_identity}.
\end{proof}

\begin{figure}
\centering
\includegraphics[width=0.6\linewidth]{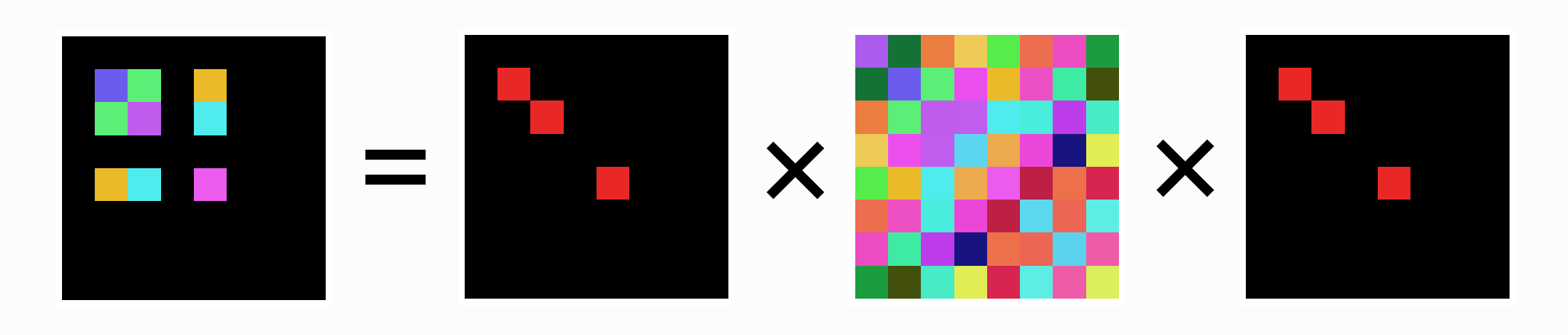}
\caption{Visualisation of the action of the projection operator $\hat{P} = \hat{P}(n, k)$ as $\hat{M}_P = \hat{P}\hat{M}\hat{P}$ for the example with $n = 3$ and $k = 1$ is shown: black squares correspond to matrix elements equal to $0$; coloured squares correspond to nonzero matrix elements; red squares correspond to matrix elements equal to $1$.}
\label{fig:projection_conjugation}
\end{figure}

The qubit projection operator $\hat{P} = \hat{P}(n, k)$ acts on the operator $\hat{M}$ to return the projected operator $\hat{M}_P = \hat{P}\hat{M}\hat{P}$,  where the original operator $\hat{M}$ and the operator following the projection $\hat{M}_P$ are $n$-qubit operator and hence can be written as $2^n \times 2^n$ matrices. All matrix elements between states that do not have Hamming weight equal to $k$ are mapped to zero following the projection. This is visualised in Fig \ref{fig:projection_conjugation}. 

We introduce results that simplify the projection of a qubit operator. Every qubit operator on $n$ qubits can be written as a linear combination of at most $4^n$ Pauli strings (tensor products of the identity operator $I$ and the Pauli operators $X$, $Y$, $Z$). As $Y = iXZ$ one can write each Pauli string as a multiplicative product of a Pauli $Z$ string (a tensor product of the identity and one-qubit $Z$ Pauli operators only) and a Pauli $X$ string. It is enough to determine the action of the projection on $X$ and $Y$ Pauli strings to determine its action on a general Pauli string. 

\begin{theorem}[Number projection of Pauli string]
The projection of a Pauli string $\hat{s} = \hat{h}_Z \hat{h}_Y$ is equal to the multiplicative product of the projection of the Pauli $Z$ string $\hat{h}_Z$ with the projection of the Pauli $Y$ string $\hat{h}_Y$ 
\begin{equation}
\begin{split}
        \hat{P}(n, k) \hat{s} \hat{P}(n, k) &= \hat{P}(n, k) \hat{h}_Z \hat{h}_Y \hat{P}(n, k)\\
        &= \hat{P}^3(n, k) \hat{h}_Z \hat{h}_Y \hat{P}(n, k)\\
        &= \hat{P}(n, k) \hat{h}_Z \hat{P}(n, k) \hat{P}(n, k) \hat{h}_Y \hat{P}(n, k
\end{split}
\end{equation}
\end{theorem}

\begin{theorem}[Number projection of Pauli Z string]
The projection by $\hat{P}(n, k)$ of a Pauli $Z$ string $H_Z$ is given by
\begin{equation}
\begin{split}
        \hat{P}(n, k) H_Z \hat{P}(n, k) &= \hat{P}(n, k)^2 H_Z\\
        &= \hat{P}(n, k) H_Z\\
        &= (n, k) H_Z \hat{P}
\end{split}
\end{equation}
\end{theorem}

\begin{theorem}[Number projection of Pauli X strings with odd terms is zero]
The projection by $\hat{P}(n, k)$ of a Pauli $X$ string that contains an odd number of $X$ one-qubit Pauli operators is equal to zero.
\begin{proof}
In order for a Pauli $X$ string to conserve the Hamming weight of some computational basis state, the number of Pauli $X$ operators it contains must be even: then the operator conserves the Hamming weight of those computational basis states that have half of the qubits on which the string does not act trivially in state $\ket{0}$ and the remaining half in state $\ket{1}$.
\end{proof}
\end{theorem}

\begin{figure}
\centering
\begin{subfigure}{0.15\textwidth}
\centering
\begin{tikzpicture}
\matrix[square matrix]
{
|[fill=lightred]| 1 & |[fill=lightred]| 1 &  \\
|[fill=lightred]| 1 & |[fill=lightcyan]| -1 &  \\
};
\end{tikzpicture}
\caption{$n = 1$}
\end{subfigure}
\hfill
\begin{subfigure}{0.2\textwidth}
\centering
\begin{tikzpicture}
\matrix[square matrix]
{
|[fill=lightred]| 1 & |[fill=lightred]| 1 & |[fill=lightred]| 1 &  \\
|[fill=lightred]| 2 & |[fill=lightgray]| 0 & |[fill=lightcyan]| -2 &  \\
|[fill=lightred]| 1 & |[fill=lightcyan]| -1 & |[fill=lightred]| 1 &  \\
};
\end{tikzpicture}
\caption{$n = 2$}
\end{subfigure}
\hfill
\begin{subfigure}{0.25\textwidth}
\centering
\begin{tikzpicture}
\matrix[square matrix]
{
|[fill=lightred]| 1 & |[fill=lightred]| 1 & |[fill=lightred]| 1 & |[fill=lightred]| 1 &  \\
|[fill=lightred]| 3 & |[fill=lightred]| 1 & |[fill=lightcyan]| -1 & |[fill=lightcyan]| -3 &  \\
|[fill=lightred]| 3 & |[fill=lightcyan]| -1 & |[fill=lightcyan]| -1 & |[fill=lightred]| 3 &  \\
|[fill=lightred]| 1 & |[fill=lightcyan]| -1 & |[fill=lightred]| 1 & |[fill=lightcyan]| -1 &  \\
};
\end{tikzpicture}
\caption{$n = 3$}
\end{subfigure}
\hfill
\begin{subfigure}{0.3\textwidth}
\centering
\begin{tikzpicture}
\matrix[square matrix]
{
|[fill=lightred]| 1 & |[fill=lightred]| 1 & |[fill=lightred]| 1 & |[fill=lightred]| 1 & |[fill=lightred]| 1 &  \\
|[fill=lightred]| 4 & |[fill=lightred]| 2 & |[fill=lightgray]| 0 & |[fill=lightcyan]| -2 & |[fill=lightcyan]| -4 &  \\
|[fill=lightred]| 6 & |[fill=lightgray]| 0 & |[fill=lightcyan]| -2 & |[fill=lightgray]| 0 & |[fill=lightred]| 6 &  \\
|[fill=lightred]| 4 & |[fill=lightcyan]| -2 & |[fill=lightgray]| 0 & |[fill=lightred]| 2 & |[fill=lightcyan]| -4 &  \\
|[fill=lightred]| 1 & |[fill=lightcyan]| -1 & |[fill=lightred]| 1 & |[fill=lightcyan]| -1 & |[fill=lightred]| 1 &  \\
};
\end{tikzpicture}
\caption{$n = 4$}
\end{subfigure}
\hfill
\begin{subfigure}{0.45\textwidth}
\centering
\begin{tikzpicture}
\matrix[square matrix]
{
|[fill=lightred]| 1 & |[fill=lightred]| 1 & |[fill=lightred]| 1 & |[fill=lightred]| 1 & |[fill=lightred]| 1 & |[fill=lightred]| 1 &  \\
|[fill=lightred]| 5 & |[fill=lightred]| 3 & |[fill=lightred]| 1 & |[fill=lightcyan]| -1 & |[fill=lightcyan]| -3 & |[fill=lightcyan]| -5 &  \\
|[fill=lightred]| 10 & |[fill=lightred]| 2 & |[fill=lightcyan]| -2 & |[fill=lightcyan]| -2 & |[fill=lightred]| 2 & |[fill=lightred]| 10 &  \\
|[fill=lightred]| 10 & |[fill=lightcyan]| -2 & |[fill=lightcyan]| -2 & |[fill=lightred]| 2 & |[fill=lightred]| 2 & |[fill=lightcyan]| -10 &  \\
|[fill=lightred]| 5 & |[fill=lightcyan]| -3 & |[fill=lightred]| 1 & |[fill=lightred]| 1 & |[fill=lightcyan]| -3 & |[fill=lightred]| 5 &  \\
|[fill=lightred]| 1 & |[fill=lightcyan]| -1 & |[fill=lightred]| 1 & |[fill=lightcyan]| -1 & |[fill=lightred]| 1 & |[fill=lightcyan]| -1 &  \\
};
\end{tikzpicture}
\caption{$n = 5$}
\end{subfigure}
\hfill
\begin{subfigure}{0.5\textwidth}
\centering
\begin{tikzpicture}
\matrix[square matrix]
{
|[fill=lightred]| 1 & |[fill=lightred]| 1 & |[fill=lightred]| 1 & |[fill=lightred]| 1 & |[fill=lightred]| 1 & |[fill=lightred]| 1 & |[fill=lightred]| 1 &  \\
|[fill=lightred]| 6 & |[fill=lightred]| 4 & |[fill=lightred]| 2 & |[fill=lightgray]| 0 & |[fill=lightcyan]| -2 & |[fill=lightcyan]| -4 & |[fill=lightcyan]| -6 &  \\
|[fill=lightred]| 15 & |[fill=lightred]| 5 & |[fill=lightcyan]| -1 & |[fill=lightcyan]| -3 & |[fill=lightcyan]| -1 & |[fill=lightred]| 5 & |[fill=lightred]| 15 &  \\
|[fill=lightred]| 20 & |[fill=lightgray]| 0 & |[fill=lightcyan]| -4 & |[fill=lightgray]| 0 & |[fill=lightred]| 4 & |[fill=lightgray]| 0 & |[fill=lightcyan]| -20 &  \\
|[fill=lightred]| 15 & |[fill=lightcyan]| -5 & |[fill=lightcyan]| -1 & |[fill=lightred]| 3 & |[fill=lightcyan]| -1 & |[fill=lightcyan]| -5 & |[fill=lightred]| 15 &  \\
|[fill=lightred]| 6 & |[fill=lightcyan]| -4 & |[fill=lightred]| 2 & |[fill=lightgray]| 0 & |[fill=lightcyan]| -2 & |[fill=lightred]| 4 & |[fill=lightcyan]| -6 &  \\
|[fill=lightred]| 1 & |[fill=lightcyan]| -1 & |[fill=lightred]| 1 & |[fill=lightcyan]| -1 & |[fill=lightred]| 1 & |[fill=lightcyan]| -1 & |[fill=lightred]| 1 &  \\
};
\end{tikzpicture}
\caption{$n = 6$}
\end{subfigure}
\hfill
\begin{subfigure}{0.45\textwidth}
\centering
\begin{tikzpicture}
\matrix[square matrix]
{
|[fill=lightred]| 1 & |[fill=lightred]| 1 & |[fill=lightred]| 1 & |[fill=lightred]| 1 & |[fill=lightred]| 1 & |[fill=lightred]| 1 & |[fill=lightred]| 1 & |[fill=lightred]| 1 &  \\
|[fill=lightred]| 7 & |[fill=lightred]| 5 & |[fill=lightred]| 3 & |[fill=lightred]| 1 & |[fill=lightcyan]| -1 & |[fill=lightcyan]| -3 & |[fill=lightcyan]| -5 & |[fill=lightcyan]| -7 &  \\
|[fill=lightred]| 21 & |[fill=lightred]| 9 & |[fill=lightred]| 1 & |[fill=lightcyan]| -3 & |[fill=lightcyan]| -3 & |[fill=lightred]| 1 & |[fill=lightred]| 9 & |[fill=lightred]| 21 &  \\
|[fill=lightred]| 35 & |[fill=lightred]| 5 & |[fill=lightcyan]| -5 & |[fill=lightcyan]| -3 & |[fill=lightred]| 3 & |[fill=lightred]| 5 & |[fill=lightcyan]| -5 & |[fill=lightcyan]| -35 &  \\
|[fill=lightred]| 35 & |[fill=lightcyan]| -5 & |[fill=lightcyan]| -5 & |[fill=lightred]| 3 & |[fill=lightred]| 3 & |[fill=lightcyan]| -5 & |[fill=lightcyan]| -5 & |[fill=lightred]| 35 &  \\
|[fill=lightred]| 21 & |[fill=lightcyan]| -9 & |[fill=lightred]| 1 & |[fill=lightred]| 3 & |[fill=lightcyan]| -3 & |[fill=lightcyan]| -1 & |[fill=lightred]| 9 & |[fill=lightcyan]| -21 &  \\
|[fill=lightred]| 7 & |[fill=lightcyan]| -5 & |[fill=lightred]| 3 & |[fill=lightcyan]| -1 & |[fill=lightcyan]| -1 & |[fill=lightred]| 3 & |[fill=lightcyan]| -5 & |[fill=lightred]| 7 &  \\
|[fill=lightred]| 1 & |[fill=lightcyan]| -1 & |[fill=lightred]| 1 & |[fill=lightcyan]| -1 & |[fill=lightred]| 1 & |[fill=lightcyan]| -1 & |[fill=lightred]| 1 & |[fill=lightcyan]| -1 &  \\
};
\end{tikzpicture}
\caption{$n = 7$}
\end{subfigure}
\hfill
\begin{subfigure}{0.50\textwidth}
\centering
\begin{tikzpicture}
\matrix[square matrix]
{
|[fill=lightred]| 1 & |[fill=lightred]| 1 & |[fill=lightred]| 1 & |[fill=lightred]| 1 & |[fill=lightred]| 1 & |[fill=lightred]| 1 & |[fill=lightred]| 1 & |[fill=lightred]| 1 & |[fill=lightred]| 1 &  \\
|[fill=lightred]| 8 & |[fill=lightred]| 6 & |[fill=lightred]| 4 & |[fill=lightred]| 2 & |[fill=lightgray]| 0 & |[fill=lightcyan]| -2 & |[fill=lightcyan]| -4 & |[fill=lightcyan]| -6 & |[fill=lightcyan]| -8 &  \\
|[fill=lightred]| 28 & |[fill=lightred]| 14 & |[fill=lightred]| 4 & |[fill=lightcyan]| -2 & |[fill=lightcyan]| -4 & |[fill=lightcyan]| -2 & |[fill=lightred]| 4 & |[fill=lightred]| 14 & |[fill=lightred]| 28 &  \\
|[fill=lightred]| 56 & |[fill=lightred]| 14 & |[fill=lightcyan]| -4 & |[fill=lightcyan]| -6 & |[fill=lightgray]| 0 & |[fill=lightred]| 6 & |[fill=lightred]| 4 & |[fill=lightcyan]| -14 & |[fill=lightcyan]| -56 &  \\
|[fill=lightred]| 70 & |[fill=lightgray]| 0 & |[fill=lightcyan]| -10 & |[fill=lightgray]| 0 & |[fill=lightred]| 6 & |[fill=lightgray]| 0 & |[fill=lightcyan]| -10 & |[fill=lightgray]| 0 & |[fill=lightred]| 70 &  \\
|[fill=lightred]| 56 & |[fill=lightcyan]| -14 & |[fill=lightcyan]| -4 & |[fill=lightred]| 6 & |[fill=lightgray]| 0 & |[fill=lightcyan]| -6 & |[fill=lightred]| 4 & |[fill=lightred]| 14 & |[fill=lightcyan]| -56 &  \\
|[fill=lightred]| 28 & |[fill=lightcyan]| -14 & |[fill=lightred]| 4 & |[fill=lightred]| 2 & |[fill=lightcyan]| -4 & |[fill=lightred]| 2 & |[fill=lightred]| 4 & |[fill=lightcyan]| -14 & |[fill=lightred]| 28 &  \\
|[fill=lightred]| 8 & |[fill=lightcyan]| -6 & |[fill=lightred]| 4 & |[fill=lightcyan]| -2 & |[fill=lightgray]| 0 & |[fill=lightred]| 2 & |[fill=lightcyan]| -4 & |[fill=lightred]| 6 & |[fill=lightcyan]| -8 &  \\
|[fill=lightred]| 1 & |[fill=lightcyan]| -1 & |[fill=lightred]| 1 & |[fill=lightcyan]| -1 & |[fill=lightred]| 1 & |[fill=lightcyan]| -1 & |[fill=lightred]| 1 & |[fill=lightcyan]| -1 & |[fill=lightred]| 1 &  \\
};
\end{tikzpicture}
\caption{$n = 8$}
\end{subfigure}
\caption{The tables show the values of $C(n, k, m)$ for $n = 1, 2, 3, 4, 5, 6, 7$ and $8$, where each row (top to bottom) corresponds to a value of $k$ and each column (left to right) corresponds to a value of $m$ (both $m$ and $k$ range from $0$ to $n$). The coefficient $C(6, 1, 2) = 5$ (third row, second column in the table for $n = 6$) can be written as the sum of coefficient $C(5, 1, 2) + C(5, 1, 1) = 2 + 3$ (third and second rows, second column in the table for $n = 5$) and alternatively as the difference $C(5, 0, 2) - C(5, 0, 1) = 10 - 5$ (third and second rows, first column in the table for $n = 5$)}.
\label{fig:generalised_binomial_coefficients_tables}
\end{figure}

\section{Numerical results}

In this section we consider the qubit Hamiltonians for example molecules of increasing complexity (in terms of both Pauli weight and number of qubits). In particular, we consider the following systems: the hydrogen molecule $\ce{H2}$; the trihydrogen cation ($\ce{H3+}$) \cite{Miller2020}; hydrogen chains of 4 and 5 atoms ($\ce{H4}$ and $\ce{H5}$); and the lithium hydride molecule $\ce{LiH}$. We use a minimal basis (STO-3G) for each example, and additionally use the double-zeta (DZ) basis for the hydrogen molecule example only. Numerical results are shown in Table \ref{table}. We observe that as expected the Pauli weight of the qubit Hamiltonian scales polynomially with the number of qubits for the Jordan-Wigner encoding, and that it scales exponentially with the number of qubits after projection; However we also evaluate the number of clique partitions \cite{Jena_2019, Gokhale2020, Kurita2023} using a greedy algorithm. Because of our choice of algorithm, the number of clique partitions we find is not guaranteed to be optimal, but it is an upper bound on the optimal value. The Pauli terms in a qubit Hamiltonian are partitioned into sets of pairwise commuting operators: the Pauli terms in each set can be simultaneously measured, significantly reducing the number of measurements needed to evaluate the energy of the Hamiltonian, notably as part of a variational quantum eigensolver (VQE) \cite{Peruzzo2014, Tilly2022} routine. The results we obtain for the number of clique partitions suggests a polynomial scaling in the number of qubits for both the Jordan-Wigner and the projected encoding. In the more computationally complex examples we consider, this means that each clique partition in the projected encoding contains a significantly larger number of terms than in the Jordan-Wigner encoding.

\begin{table}[h!]
\centering
\begin{tabular}{cllrrrrr}
\toprule
 & Molecule   & Basis set   &   Qubits &  &  Pauli weight &    &  Clique partitions \\
 \cmidrule(r){5-6} \cmidrule(r){7-8}
 &&&&Jordan-Wigner&Projected&Jordan-Wigner&Projected\\
\midrule
 \adjustbox{valign=c}{\includegraphics[width=0.7 cm, height=0.7cm]{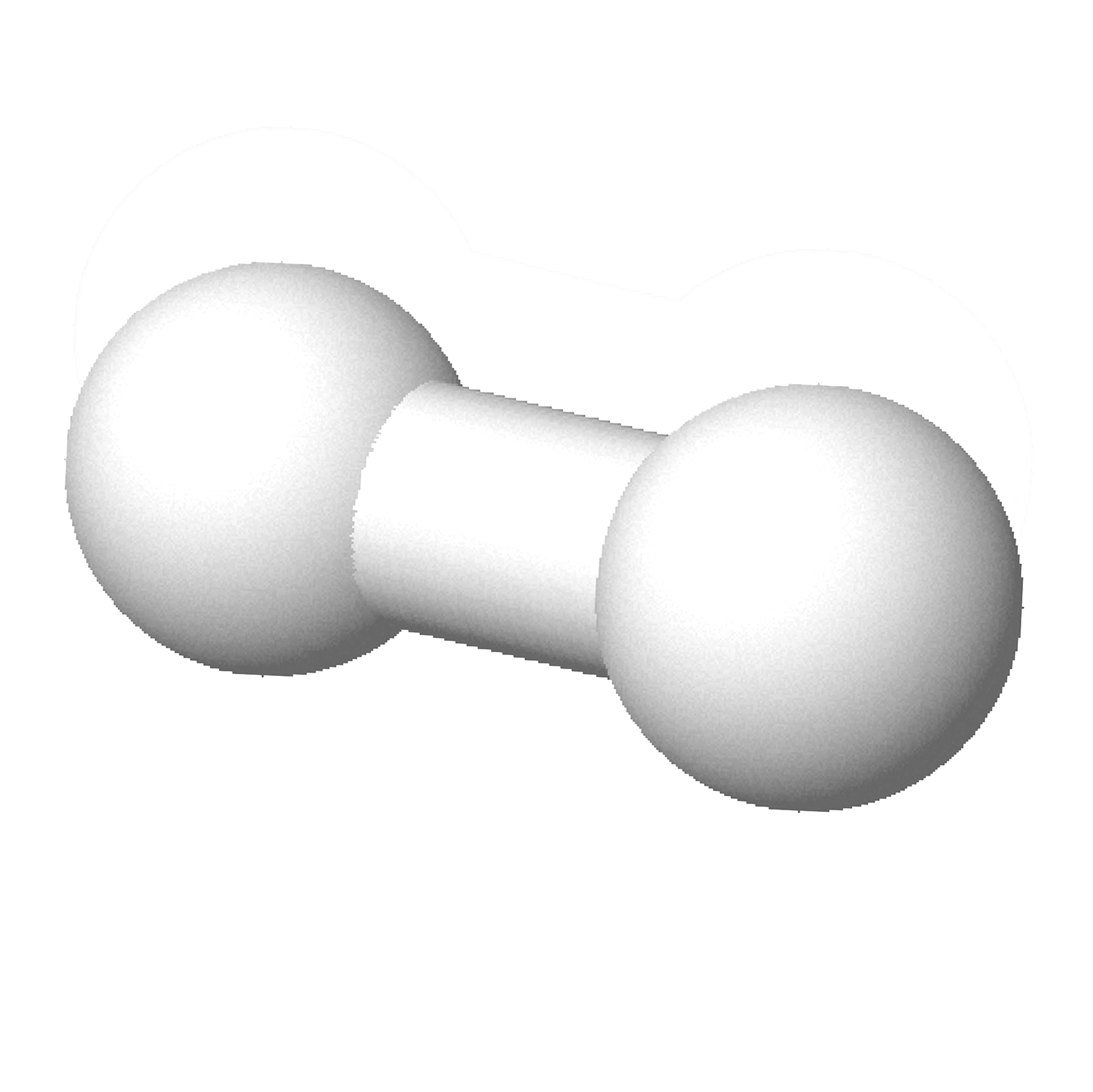}}&Hydrogen molecule ($\ce{H2}$)         & STO-3G  &        4 &                  15 &                          20 &               2 &                      2 \\
 \adjustbox{valign=c}{\includegraphics[width=0.8cm, height=0.8cm]{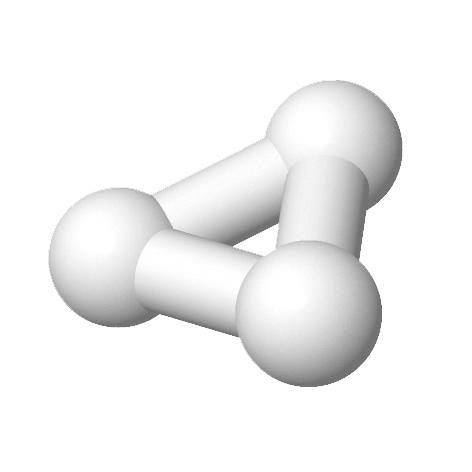}}&Trihydrogen cation ($\ce{H3+}$)        & STO-3G  &        6 &                  62 &                         184 &               8 &                      8 \\
 \adjustbox{valign=c}{\includegraphics[width=0.7cm, height=0.7cm]{figures/H2.png}}&Hydrogen molecule ($\ce{H2}$)         & DZ      &        8 &                 185 &                        3,454 &              10 &                     27 \\
 \adjustbox{valign=c}{\includegraphics[width=1.2cm, height=0.6cm]{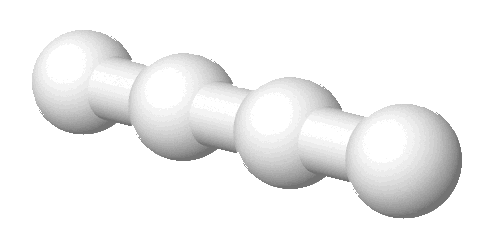}}&Hydrogen chain ($\ce{H4}$)         & STO-3G  &        8 &                 185 &                        1,948 &              10 &                     16 \\
 \adjustbox{valign=c}{\includegraphics[width=1.4cm, height=0.7cm]{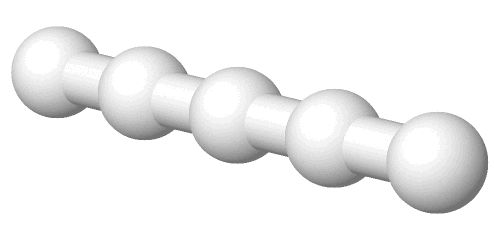}}&Hydrogen chain ($\ce{H5}$)         & STO-3G  &       10 &                 436 &                       17,744 &              33 &                     64 \\
 \adjustbox{valign=c}{\includegraphics[width=1cm, height=1cm]{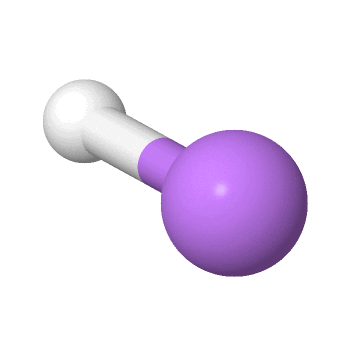}}&Lithium hydride ($\ce{LiH}$)        & STO-3G  &       12 & 631 &                      155,300 &              41 &                     84 \\
\bottomrule
\end{tabular}
\caption{Numerical results for example molecules, where the Pauli weight is the number of terms in the decomposition of the Hamiltonian in terms of Pauli operators, and clique partitions is an upper bound on the number of commuting clique partitions.}
\label{table}
\end{table}

\section{Conclusion}

We have discussed the explicit construction of qubit number projection operators and their relationship to Kravchuk coefficients, visualized through Pascal's pyramid. We have provided proof for several  properties of these coefficients, such as their recursive relations and orthogonality. Qubit number projection operators are important for many-body quantum systems where number conservation symmetry arises; in particular, we have proposed one application to variational quantum alogrithms for quantum chemistry, and considered the complexity of projected qubit Hamiltonians for a number of example molecular systems. We are actively working on further areas of research, which include employing the properties of the projection operators we have discussed to develop more efficient algorithms for enforcing particle number conservation in quantum computations.

\section*{Acknowledgements}

D.P.'s research is supported by an industrial CASE (iCASE) studentship, funded by the Engineering and Physical Sciences Research Council (EPSRC) [EP/T517793/1], in collaboration with University College London and Rahko Ltd. D.P. would like to thank Jonathan Tennyson for useful discussions.

\bibliography{bibliography}

\end{document}